\documentclass [12pt,a4paper] {article}
\title {A SIMPLE HOLOGRAPHIC MODEL OF THE COSMOLOGICAL CONSTANT}

\author {Vladan Pankovic, Simo Ciganovic$^\sharp$\\
Jovan Ivanovic$^\sharp$, Rade Glavatovic$^\diamond$, Petar Grujic
$^\S$\\$^\ast$Department of Physics, Faculty of Sciences, \\
21000 Novi Sad,Trg Dositeja Obradovica 4, Serbia, \\
$^\sharp$Gimnazija, 22320 Indjija, Trg Slobode 2a, Serbia \\
$^\diamond$ Military-Medical Academy, 11000 Belgrade, Crnotravska
17, Serbia \\
$^\$$ Institute of Physics, P.O. Box 57, 11000 Belgrade, Serbia
}

\date{}

\begin {document}

\maketitle 
\vspace {.5cm}


\vspace {.5cm}

\begin {abstract}
We examine a simple theoretical model to estimate (by fine tuning
condition) the value of the cosmological constant. We assume, in
analogy with holographic principle, that cosmological constant, like
classical surface tension coefficient in a liquid drop, does not
correspond to a volume (bulk) vacuum mass (energy) density
distribution, but rather to the surface vacuum mass (energy) density
distribution. Then the form of given surface mass distribution and
fine tuning condition imply observed growing (for about 61 order of
magnitude) of the scale factor, from the initial (corresponding to
Planck length), to the recent, at the beginning of the cosmic
acceleration (corresponding to 10 Glyr length).
\end {abstract}
\vspace {1.cm}

 \section{Introduction}

       Modern cosmology introduces, from one side, new ingredients into  the overall
       content of the Universe (ontological aspects) and parallelly
       contrives new theoretical models (or adopts them from other
       fields), from the methodological side. Dark energy assumption
       has been one of the principal component of the newly
       conceived Universe, though it has net yet been clarified as
       for the physical content of this hypothetical quantity [1,3].
       At present, there are three candidates for the physical
       realization of this component, which is believed to comprise
       about 70 per cent of the overall mass-energy content of the universe
       (se,e.g. [3]): (i) {\it quintessence}, (ii) {\it cosmological constant},
       (iii) {\it phantom fields}. As is well known, Einstein's cosmological
       constant $\Lambda$ was introduced in order to keep the GR Universe stable,
       but we stress here that the concept of the cosmological constant goes beyond
       the general relativity and may be introduced via Newtonian classical dynamics
       (see, e.g. [2]).As for the modeling the structure and evolution of the Universe, many
       proposals appear on the market, with ever increasing number.
       One of the prominent concepts, in this context, has been the
       so-called holographic paradigm.

           The latter had previous history before appearing
           in astrophysics and cosmology. The concept of determining
           the properties of a manifold by those of a related
           submanifold, appears in many integral theorems,like
           Gauss'and Stock's ones. The astrophysics of black holes has
           renewed the interest in the holographic paradigm, with
           Beckenstein-Hawking determination of the black hole
           entropy. Since many cosmological models have been based
           on the concept of black hole, it is not surprising that
           some properties of the latter has served as an ansatz of the
           cosmological models. Recently it has been argued that the
           observed cosmic acceleration may be explained by the
           interaction between the pressureless dark matter and
           holographic dark energy [4], with cutoff scale determined
           by the Hubble length. Further, gravitational holography
           is believed to render stability to the cosmological
           constant, via quantum contributions to the stress-energy
           tensor in the expanding Universe [5].

             On the more abstract side, cosmological models often appear
             construed according to modeling in various fields of
             physics. It holds for the Beckenstein-Hawking black hole entropy,
             which was inspired by the solid state physics [4].
             In fact, many cosmological models have been inspired by
             the classical laboratory (Earthly) phenomena, albeit
             implicitly. Methodologically, such an approach
             compromises possible cosmic reality with the
             transparency of the physical, classical models, as
             useful approximations to the cosmic features. It is
             this rationale we are making use in determining the
             value of the cosmological constant, as we are going to
             expose in the following.

             \section{The model and results}

      We start with Friedmann equation (see for example [6]), without cosmological
constant term
\begin {equation}
   \Bigl(\frac {(\frac {d{\it a}}{dt})}{{\it a}}\Bigr)^{2} + \frac {kc^{2}}{{\it a}^{2}}=
    G(\frac {8\pi}{3})\rho
\end {equation}
(where  ${\it a}$ represents the  scale factor of the universe, $k$
– curvature constant (that equals $1$ for closed, $0$ for flat and
$-1$ for open universe), $c$ – speed of light,
 $G$ – Newtonian gravitational constant, $\rho$ – total mass density) can be formally
 interpreted by classical, Newtonian mechanics and gravitation in the following way.
Namely, equation (1) can be equivalently, simply transformed in the following equation
\begin {equation}
   \frac {m(dr/dt)^{2}}{2} + \frac {kmc^{2}}{2}= mG(\frac {4\pi}{3})r^{3}\frac {\rho}{r}
\end {equation}
where $r={\it a}{r_{0}}$ represents sphere radius, $r_{0}$ -
appropriate length unit, while $m$ can be considered as the mass of
a classical probe system, a particle or spherical shell. Term $ m
(dr/dt)^{2}/2$ corresponds then to the classical kinetic energy, and
the constant term $kmc^{2}/2$ can be formally taken to be the total
energy of a classical harmonic linear oscillator or rotator.
Finally, term $ mG(4\pi/3)r^{3}(\rho/r)$ can be formally interpreted
as the classical potential energy by gravitational interaction
between probe system and universe with mass $(4\pi/3)r^{3}\rho$
homogeneously distributed with density $\rho$ within the sphere
volume $(4\pi/3)r^{3}$.

Friedmann equation with additional cosmological constant term
\begin {equation}
   \Bigl(\frac {da/dt}{a}\Bigr)^{2} + \frac {kc^{2}}{{\it a}^{2}}=
   G(\frac {8\pi}{3})\rho + \frac {\Lambda}{3}
\end {equation}
(where $\Lambda$ represents the cosmological constant) can be, also,
formally interpreted classically. Namely, it can be equally
transformed into the following equation

\begin{equation}
 {m(dr/dt)^2\over 2} + {kmc^2\over 2} = mG{(4\pi/3)r^3\over r} + m{\Lambda\over24\pi}(4\pi r^2)
\end {equation}

Additional term $ m\Lambda/(24 \pi) (4\pi r^{2}) $ can formally be
considered as the classical energy of the surface tension of a fluid
captured in the sphere with radius $r$ and surface area $(4\pi
r^{2}) $ so that surface tension coefficient is $ m\Lambda/24 \pi$.
It is important to point out that here (as well as in the fluid
mechanics), even if surface tension coefficient is constant, energy
of the surface tension increases when sphere radius increases.

  We note here that the surface tension of the fluid captured in the sphere,
   i.e. of a liquid drop (corresponding to material universe), is directed radially,
   from sphere surface toward sphere center, similar to the gravitational tension
   of a fluid. However, this surface tension is not directed in the opposite direction,
   radially from sphere center toward sphere surface, as it is necessary for an
   anti-gravitational, vacuum action. Nevertheless, we can imagine a spherical bubble
   of an ideal gas (corresponding, here, to the material universe) imbedded in the bulk
   of a fluid, i.e. liquid (corresponding here to the dark energy vacuum [11]).
    In this case surface tension (corresponding to the vacuum action) of the liquid without spherical bubble is
    directed radially from sphere center toward sphere surface, in other words
    "anti-gravitationally".

All this implies a possibility [7] that cosmological constant, not
only formally classically, but even really corresponds to some
surface phenomena. (Especially, it can be observed that all this
corresponds, at least conceptually, to t'Hooft-Susskind holographic
principle in quantum gravity [8].) One aspect of this possibility we
shall briefly consider here. Namely, we consider formulating a
satisfactory, simplified theoretical prediction of the data
corresponding to observationally (by fine tuning condition)
estimated value of the cosmological constant [9],[10].

   We assume that cosmological constant, like surface tension
coefficient by a liquid drop, does not correspond to a volume (bulk)
vacuum mass (energy) density distribution but that it corresponds to
a surface vacuum mass (energy) density distribution. Specifically,
we suppose that vacuum mass is distributed over a thin spherical
shell with sphere radius proportional to scale factor and thickness
equivalent to Planck length. Then form of given surface mass
distribution and fine tuning condition imply observed value of the
scale factor, i.e. growing of the scale factor (for approximately 61
order of magnitude) from the initial (corresponding to Planck length
about $10^{-35}m$) to the recent (at the beginning of the cosmic
acceleration, corresponding to $10 Glyr \sim 10^{26}m$ length).

So, we take, as usually,
\begin {equation}
   G(\frac{8\pi}{3})\rho_{\Lambda} = \frac {\Lambda}{3}
\end {equation}
where $\rho_{\Lambda}$ represents the mass density corresponding to cosmological constant.

Further, we suppose
\begin {equation}
  \rho_{\Lambda} = \frac {M_{\Lambda}}{L_{P}4\pi r^{2}}
\end {equation}
which corresponds to vacuum mass $ M_{\Lambda}$ homogeneously distributed over thin
spherical shell with sphere radius $r$ and thickness equivalent to Planck length $ L_{P}$.
It implies
\begin {equation}
  M_{\Lambda}  = \rho_{\Lambda}( L_{P}4\pi) r^{2}
\end {equation}
which means that vacuum mass grows quadratically with the sphere
radius.

Accounting for(6) and inserting
\begin {equation}
  \Lambda = \frac {c^{2}}{L^{2}_{\Lambda}}
\end {equation}
(where $L_{\Lambda}$ represents the length corresponding to cosmological constant)
 in (5), after simple transformations, yields
\begin {equation}
  \frac {M_{\Lambda}}{M_{P}}  = \frac {1}{2} \frac {r^{2}}{ L^{2}_{\Lambda}}                      .
\end {equation}
We obtain an interesting result allowing for comparison with the
observational data.

Initially, i.e. for $r = L_{P}$, and, according to fine-tuning condition [4], [5]
\begin {equation}
     \frac { L^{2}_{P}}{ L^{2}_{\Lambda}} \sim 10^{- 123}                ,
\end {equation}
it follows
\begin {equation}
  \frac {M_{\Lambda}}{M_{P}}= \frac {1}{2} \frac { L^{2}_{P}}{ L^{2}_{\Lambda}}
\end {equation}
or
\begin {equation}
  M_{\Lambda} \sim 10^{- 123} M_{P}         .
\end {equation}
Evidently, it means that cosmological constant does not have any
important influence in the early universe.

Now, we shall determine by (9) such $r$ for which condition
\begin {equation}
    M_{\Lambda}\sim M_{P}
\end {equation}
is satisfied. Given condition, of course, simply means that vacuum energy becomes
comparable with energy of the quantum fields. Introduction of (13) in (9) yields,
after simple transformations,
\begin {equation}
   r^{2} \sim  L^{2}_{\Lambda}      .
\end {equation}
Now, we shall express $r$ in the following way
\begin {equation}
   r = {\it a} L_{P}
\end {equation}
which, introduced in (14), yields
\begin {equation}
  {\it a}^{2} \sim \frac { L^{2}_{\Lambda}}{ L^{2}_{P}} \sim 10^{123}
\end {equation}
and further
\begin {equation}
  {\it a} \sim 10^{61}                   .
\end {equation}
This appears a remarkable result. Namely, it corresponds
satisfactorily to the observational data [9],[10] on the growing of
the scale factor of the universe for about $61$ order of magnitude,
from the initial (corresponding to Planck length approximately
$10^{-35}m$) to the recent (at the beginning of the cosmic
acceleration, corresponding to about $10 Glyr \sim 10^{26}m$
length).

   We note here that, as it is not difficult to see, the assumption about the
volume, i.e. bulk distribution of the vacuum mass yields
unsatisfactory prediction ${\it a} \sim 10^{41}$.

\section{Conclusion}

     In conclusion we point out that suggested model of the interpretation of
     cosmological constant as a surface phenomena (at least conceptually analogous
     to t'Hooft-Susskind holographic principle) is very rough and simplified at present,
     formally looking like a classical one. Such model must be necessarily generalized,
     but it goes beyond the basic intention of our work. In any case our model is able to
     correlate and reproduce observational astronomical data (fine-tuning and growing
     of the scale factor) in a satisfactory way. We consider this an interesting and
     promising result.

\section {References}

\begin {itemize}
\item [[1]] H-J. Fahr and M. Heyl, {\it Naturwissenschaften}, DOI 10.
1007/s00114-007-0235-1
\item [[2]] J. W. Norbury, {\it Eur. J. Phys.} {\bf 19} (1998)143-150.
\item [[3]] J. A. de Freitas Pacheco and J. E. Horvath, {\it Class.
Quantum Grav.} {\bf 24} (2007) 5427-5433.
\item [[4]] W. Zimdahl and D. Pavon, {\it Class. Quantum Grav.} {\bf 24} (2007) 5461-5478.
\item [[5]] S. Thomas, {\it Phys. Rev. Lett.} {\bf 89} (2002) 081301.
\item [[6]] B. W. Carroll, D. A. Ostlie, {\it An Introduction to Modern Galactic Astrophysics and Cosmology} (Addison-Wesley, Reading, MA, 2007)
\item [[7]] T. Padmanabhan, {\it Dark Energy: Mystery of the Millennium}, astro-ph/0603114 and references therein
\item [[8]] R. Bousso, {\it The holographic principle}, hep-th/0203101 and references therein
\item [[9]] D. N. Spergel et al., Astrophys. J. Supp. {\bf 146} (2003) 175 ; astro-ph/0302209.
\item [[10]] D. N. Spergel et al., Astrophys. J. Supp. {\bf 170} (2007) 337 ; astro-ph/0603449.
\item [[11]] P. Grujic, {\it Astrophys. Space Sc.}, {\bf 295} (2004) 363-374.

\end {itemize}

\end {document}